\documentstyle[aps,twocolumn]{revtex}
\input epsf
\tightenlines
\begin{document}
\draft
\preprint{VECC-2000-}
\title{
 Radiation of single photons from $Pb+Pb$ collisions at the CERN
SPS and quark hadron phase transition}
\author{Dinesh Kumar Srivastava$^1$ and Bikash Sinha$^{1,2}$}
\address{$^1$Variable Energy Cyclotron Centre, 1/AF Bidhan Nagar, Calcutta 
700 064\\
$^2$Saha Institute of Nuclear Physics, 1/AF Bidhan Nagar, Calcutta
700 064 }
\date{\today}
\maketitle

\begin{abstract}
The production of single photons in $Pb+Pb$ collisions
at the CERN SPS as measured by the WA98 experiment is analysed.
A very good description of the data is obtained if a
quark gluon plasma is assumed to be formed initially, which expands,
cools, hadronizes, and undergoes  freeze-out.
A rich hadronic equation of state is used and the transverse expansion of the
interacting system is taken into account. The recent estimates of photon
production in quark-matter (at two loop level) along with the dominant
reactions in the hadronic matter leading to photons
are used. About 50\% of the single photons are seen to have a 
thermal origin. An addition of the thermal and prompt
photons is seen to provide a very good description of the data.
Most of the thermal photons  having large transverse momenta
arise from the quark-matter, which contributes dominantly through
the mechanism of annihilation of quarks with scattering, and which in turn
is possible only in a hot and dense plasma of quarks and gluons.
The results thus confirm the formation of quark gluon plasma
and the existence of this mechanism of the production of single
photons.

\end{abstract}
\narrowtext

The search for quark-gluon plasma, which filled the early
universe microseconds after the big bang and which may be present in the
 core of neutron stars, is one of the most notable collective efforts
 of the present day nuclear physics community. 
Its discovery will provide an important confirmation of the
predictions of the statistical Quantum Chromodynamics (QCD) based on
 lattice calculations. It has been
recognised for a long time~\cite{early}  that electromagnetic radiations from 
relativistic heavy ion collisions in these experiments would be 
a definitive signature of the
formation of a hot and dense plasma of quarks and gluons, consequent to a
quark-hadron phase transition~\cite{early}.
Once other signs of the quark-hadron transition, e.g., 
an enhanced production of strangeness, a suppression of $J/\psi$
production,  radiation of dileptons,
etc., started to emerge~\cite{qm99}, it was imperative that
the more direct, yet much more difficult to isolate, signature
of the hot and dense quark-gluon plasma, the single photons
were identified. The WA98 experiment~\cite{wa98} has now reported
observation of single photons in central $Pb+Pb$ collisions at the
CERN SPS.

In the present work we show that these data are
very well described if we assume that a quark-gluon plasma was
formed in the collision.

In order to put our findings in a proper perspective, 
let us  recall that the publication of the upper limit of the production of
single photons in $S+Au$ collisions at CERN SPS~\cite{wa80}
by the WA80 experiment was preceded and followed by 
several papers~\cite{prl,others}
exploring their connection to the quark-hadron phase
transition. An early work, by the present authors~\cite{prl}, reported
that the data were consistent with a scenario where a quark gluon
plasma was formed at  an initial time $\tau_0\sim$ 1 fm/$c$, which expanded
and cooled, got into a mixed phase of quarks, gluons, and hadrons,
and ultimately underwent a freeze-out  from a  state of hadronic gas
consisting of $\pi$, $\rho$, $\omega$, and $\eta$ mesons. 
On the other hand, when the initial state was assumed to consist
of (the same) hadrons, the resulting large initial temperature led to a
much larger production of single photons, in  gross violation of the
upper limit.

A reanalysis of the WA80 data on single photons was reported 
recently~\cite{zpc1} which incorporated two important developments in 
the field during the last few years, which are worth recalling.
 Firstly, it was realized 
that the hadronic equation of state {\em must} be generalized to
include all of the hadrons~\cite{crs} 
 ( limited to $M <$ 2.5 GeV, in practice). This was prompted and
supported by the success of the
thermal models in describing particle production in these 
collisions. This implied that the hadrons were in
chemical equilibrium~\cite{johanna} at least at the time of
 (chemical) freeze-out. These hydrodynamical calculations have been 
shown to provide a very good explanation of the $p_T$ spectra 
measured by the NA49 and NA44 experiments~\cite{had_dil}. 

Secondly, an evaluation of the rate of single photon production from the
quark matter to the order of two-loops was reported recently
by Aurenche et al~\cite{pat,MT}. This had two quite important results:
 (i) a substantial contribution of the bremsstrahlung
($q\,q\,(g)\,\rightarrow\,q\,q\,(g)\,\gamma$) process 
for all momenta in addition to the Compton 
($q\,(\overline{q})\,g\,\rightarrow\,q\,(\overline{q})\,\gamma$) plus
 annihilation ($q\,\overline{q}\,\rightarrow\,g\,\gamma$)
contributions included in the one-loop calculations available in the
literature~\cite{joe,rolf}, and (ii) a large contribution by a new mechanism
which corresponds to the annihilation of a quark (scattered from a
quark or a gluon) by an anti-quark.
These new rates were shown~\cite{zpc2} to  lead to  a considerable
 enhancement of the
production of single photons at SPS, RHIC, and LHC energies,
if the initial state is approximated as an equilibrated plasma.

It was also reported~\cite{zpc1} that when allowances were made for the above
considerations, the WA80 upper limit was still consistent with a quark
hadron phase transition, while a treatment without phase transition
 was untenable as it involved several hadrons/fm$^3$, at the initial time.

We add that there can be a production of high momentum
single photons during the pre-equilibrium phase, when treated within the
parton cascade model~\cite{pcm}, from the fragmentation of time-like 
quarks ($q\,\rightarrow\,q\,\gamma$) produced
in (semi)hard multiple scatterings~\cite{pcmphot}.

The rate for the production of hard photons evaluated to one
loop order using the effective theory based on resummation of
hard thermal loops is given by~\cite{joe,rolf}:
\begin{equation}
E\frac{dN}{d^4x\,d^3k}=\frac{1}{2\pi^2}\,\alpha\alpha_s\,
                          \left(\sum_f e_f^2\right)\, T^2\,
                       e^{-E/T}\,\ln(\frac {cE}{\alpha_s T})
\end{equation}
where the constant $c\approx$ 0.23.  The summation runs over
the flavours of the quarks and $e_f$ is their electric charge 
in units of charge of the electron. The rate of production
of photons due to the bremsstrahlung processes evaluated by
Aurenche et al is given by:
\begin{eqnarray}
E\frac{dN}{d^4x\,d^3k}&=&\frac{8}{\pi^5}\,\alpha\alpha_s\,
                          \left(\sum_f e_f^2\right)\, 
                        \frac{T^4}{E^2}\,\times \nonumber\\
                       & &e^{-E/T}\,(J_T-J_L)\,I(E,T),
\end{eqnarray}
and the expressions for $J_T$, $J_L$, and $I(E,T)$ can be found 
in Ref.~\cite{pat}.

 And finally the dominant contribution of the $q\overline{q}$ annihilation
with scattering obtained by Aurenche et al is given by:
\begin{equation}
E\frac{dN}{d^4x\,d^3k}=\frac{8}{3\pi^5}\,\alpha\alpha_s\,
                          \left(\sum_f e_f^2\right)\, ET \,
                          e^{-E/T}\,(J_T-J_L).
\end{equation}
Note that all the three contributions turn out to be  essentially of the order
$\alpha \alpha_s$ \cite{pat}. {\em It has been pointed out recently
~\cite{MT} that the
values of $J_T$ and $J_L$ given originally by Aurenche et al~\cite{pat}
 are too large by a numerical factor of 4. We use the corrected values in the
following.}
 
The estimate of prompt photons is taken from the work of Wong and 
Wang~\cite{wong} which employs the NLO pQCD along with the inclusion of the
effects of intrinsic partonic momenta ($ <k_T^2>=$ 0.9 GeV$^2$;
see discussion later).

We assume that a chemically and thermally equilibrated quark-gluon
plasma is produced in such collisions at the time $\tau_0$ (see
later), and use the isentropy condition~\cite{bj};
\begin{equation}
\frac{2\pi^4}{45\zeta(3)}\,\frac{1}{A_T}\frac{dN}{dy}=4 a
T_0^3\tau_0
\label{T0}
\end{equation}
to estimate the initial temperature, where $A_T$ is the
transverse area. 

We have
taken the  average particle rapidity density as 750 for the 10\% most
 central $Pb+Pb$ collisions at the CERN SPS energy
as measured in the experiment.  We estimate the average number of
participants for the corresponding range of impact parameters
($0\, \leq\, b \, \leq \, 4.5$ fm) as  about 380, compared to the maximum of
416 for  a head-on collision. We thus use a mass number of 190 to get the
 radius of the
transverse area of the colliding system and neglect
its deviations from azimuthal symmetry, for simplicity.
As this deviation, measured in terms of the number of participants, is
marginal ($<$ 9\%) we expect the error involved to be small. We also
recall that the azimuthal flow is minimal for central collisions.

 We take $a=42.25\pi^2/90$ for
a plasma of massless quarks (u, d, and s) and gluons, where we have 
put the number of flavours as
$\approx$ 2.5 to account for the mass of the strange quarks.
We now use Eq.(\ref{T0}) to estimate the (average) initial temperature, 
with the  
additional assumption of a rapid thermalization~\cite{kms} so that
the formation time is decided by the uncertainty relation and $\tau_0=1/3T_0$.
This $T_0$ is then used to get the (average) initial energy density.

It is important to have a proper initial energy density profile
as it affects the hydrodynamic developments by introducing additional gradients.
We assume it to follow the so-called
`wounded-nucleon' distribution, which for central collision of identical
nuclei leads to:
\begin{equation}
\epsilon(\tau_0,r)\propto \int_{-\infty}^\infty \rho(\sqrt{r^2+z^2})\, dz
\end{equation}
where $\rho$ is the (Woods-Saxon) distribution of nucleons in a nucleus
having  a mass number  of $190$
and $r$ is the transverse distance. This is prompted by
the experimental observation that transverse energy deposited in these
collisions scales with the number of participants.  The normalization in the
above is
determined from a numerical integration so that:
\begin{equation}
A_T \, \epsilon_0=\int \, 2\pi \, r \,  \epsilon(r) \,\,dr.
\end{equation}

We further assume that the phase transition takes place at $T=$ 180 MeV and the
freeze-out takes place at 120 MeV. This value of the critical temperature
is motivated by the recent lattice QCD results which give values of about 
170 -- 190 MeV~\cite{kars}, and the thermal model analyses of hadronic 
ratios which 
suggest that the chemical freeze-out in such collisions takes place
at about 170 MeV. ( A recent analysis by Becattini et al yields a value
 of $181.3\,\pm\,10.3$ MeV~\cite{johanna} for the chemical freeze-out
temperature.)
The phase transition should necessarily take place at a higher
temperature. 

The rates for the hadronic matter have been 
obtained~\cite{joe}
from a two loop approximation of the photon self energy 
using a model where $\pi-\rho$ interactions have been included. The 
contribution of the $A_1$ resonance is also included according to the
suggestions of Xiong et al~\cite{li}. The relevant hydrodynamic equations are
solved using the procedure~\cite{hydro} discussed earlier and
an integration over history of evolution is performed~\cite{crs}.

In Fig.~1 we show our results.  The dashed curve gives the
contribution of the quark-matter and the solid curve gives the
sum of the contributions of the quark matter and the hadronic matter.
The NLO pQCD estimates for prompt photons $pp$ are also given.
 We see that the thermal photons  contribute to about 50\% of the total
yield of the single photons and that the sum of thermal and prompt 
photons provides a very good description to the data. We also note that
at higher transverse momenta most of the thermal photons have their
origin in the quark matter.

How sensitive are the results to the choice of our parameters?
In Fig.~2, we show our results where we vary the transition temperature
by $\pm$ 20 MeV. It is seen that the results at higher $k_T$
(which have their origin in earlier times) remain unaltered, though the
yield at the lowest transverse momenta increases with the decrease in
$T_C$.

The initial time $\tau_0$ affects the results much more strongly, as
increasing it lowers the initial temperature (Eq.\ref{T0}). In Fig.~3(a) we show
our results for $\tau_0=$ 0.20, 0.40, 0.60, 0.80 and 1 fm/$c$, corresponding to
$T_0=$ 335, 265, 232, 210, and 196  MeV.  A comparison of this figure with
 Fig.~1 shows that the data  clearly favour
a large initial temperature (and early thermalization). Recall that the
hydrodynamic flow of all the systems (having same $dN/dy\sim T_0^3 \tau_0$)
 are known to be nearly identical at later
times~\cite{hydro} and thus affect the hadronic 
data only marginally; see Fig.3(b).

A very important outcome of these results (Fig.~1) is that a very 
large part of the thermal component
of the single photons is seen to have its origin in the quark-matter
itself!
Recall that the new (and dominant) mechanism of the annihilation of
quarks with scattering, suggested by Aurenche et al., is operative
{\em only} if a hot and dense plasma is formed (see the 
detailed discussion in the Appendix in Ref.~\cite{pat}).
Thus these results confirm the existence of this mechanism and the 
formation of quark gluon plasma in such collisions.

Even though we realize that the creation of a hot (confined)
hadronic matter in thermal 
and chemical equilibrium within $\tau_0\approx$ 0.20 fm/$c$, consequent
to nuclear collision is highly unlikely~\cite{kap}, we estimate the
initial temperature for such a system  from Eq.(\ref{T0}) 
for the hadronic equation of state used here, as more than
260 MeV, when the hadronic density would be $\approx$ 10
 hadrons/fm$^3$~\cite{crs}. We consider this very unphysical and unlikely.
A larger formation time will  give  a much lower initial
temperature and fail to explain the data.

How are we to understand the
use of $\tau_0 =1/3T_0\approx $ 0.20 fm/$c$ here (see also~\cite{kms}) against 
the  canonical value of 1 fm/$c$, employed often? Firstly, 
within the model used, this value is {\em favoured} by the data (Fig.~3).
  Secondly, if a larger value
of $\tau_0$ is used, then an allowance should be made to supplement the
predictions with an appropriate pre-equilibrium contribution
(see e.g., Ref.~\cite{pcmphot2}).
Thirdly, we note that the matter at $z=0$ 
starts interacting by $t=-R/\gamma\approx\, -0.7$ fm/$c$ in the
present case, when the two nuclei start touching. Thus by the lapse of
$\tau=0.2$ fm/$c$, the matter there has been under interaction 
for a time $\sim$ 1 fm/$c$, which may be enough for the formation 
of the plasma.

Finally, a very important confirmation of our findings comes from
the observations of Eskola et al~\cite{sat}, that
that a saturation of partons signaling
a complete filling up of the transverse area  by coloured quanta
in collision of lead nuclei at SPS energies is indeed attained when
the momentum transfer in partonic collisions is of the order of 1 GeV
leading to a temperature $\sim$ 300 MeV at $\tau_0\sim$
0.2 fm/$c$.

We add that the model developed here provides a very good
description~\cite{had_dil} of the intermediate mass dilepton 
excess measured by the NA50 group. 

Are we justified in making the assumption of a chemically equilibrated plasma,
 considering that
indeed the predictions at the lower transverse momenta are close to the
upper limits given by the experiment? This needs to be investigated (see
Neumann et al~\cite{others}) as also
the effect of (likely) medium modification of hadron properties.
The neglect of the baryo-chemical potential for the QGP is perhaps 
justified as the net-baryon to hadron ratio is quite small~\cite{xu},
 especially in the region of the central rapidity. 
Finally, we may add that the photon rates used in these calculations
are strictly valid only for $\alpha_s \ll$ 1 and  that the consequences
of considering higher loops remains to be seen.

Before summarizing, let us return to the question of prompt photons.
A detailed discussion on them is beyond the scope of this paper,
and the debate on the reproducibility of single photons data in
fixed target  $pp(A)$ experiments is inconclusive.
  If we are to believe the results of Wong and Wang~\cite{wong},
which we have employed, then the prompt photons contribute  
about half of the total yield in the present work,
As mentioned earlier, these results are
obtained by using NLO pQCD predictions along with the inclusion of 
intrinsic momenta of partons.

  The classic paper of Owens~\cite{owens} discusses
the need to account for the intrinsic transverse momenta of partons. However,
that work also talks of the need to introduce a cut-off in $Q^2$,
below which pQCD can not be applied and to avoid singularities
in the parton-parton matrix elements. This discussion is absent
in Ref.\cite{wong,jane}.  Large 
enhancements can be obtained depending on the the cut-offs employed
and the $<k_T>$ used. The extent to which these considerations
will affect the results of Wong and Wang is not known.

We also recall the exhaustive work of 
Vogelsang and Whalley~\cite{pqcd}; especially their Fig. 31,
where the differences of {\em all} the pp data from NLO pQCD predictions are 
plotted.  It is seen that while all the pp data for $\sqrt{s} >$ 23 GeV
are quantitatively explained by  NLO pQCD, the one at 19.4 GeV
 is underpredicted by a factor of 4--5. If this trend is assumed 
to continue then at the relevant nucleon-nucleon energy of 17.3 GeV,
this difference would mount to a factor $>$ 10 (!). The NLO pQCD
analysis of these authors is  also applied to $p+Be$ data at 31 GeV
(by normalizing it to $pp$) and a very good description is obtained,
while the same procedure underestimates the $p+C$ data at 19.4 GeV
by a factor of 4--5. Several of these data have also been critically
examined by Aurenche et al.~\cite{aur2} within NLO pQCD, 
who conclude that: i) there
is no need to include intrinsic momentum effects, ii) the perturbation
theory becomes unstable at lower $k_T$ when intrinsic momenta of partons
is included, and iii) the data at lower energies are 
incompatible with those at higher energies, especially the $p+Be$
data. Several other papers (see Ref.~\cite{aur2}) also discuss
these aspects. 

On the other hand authors of Ref.~\cite{papp} have studied the effect
of the so-called $p_T$ broadening (Cronin effect) in 
proton-nucleus collisions. The LO pQCD is used
along with  a $K$ factor and an intrinsic $<k_T>$ for the partons. 
The fact that the $p+Be$ data is explained reasonably well by Vogelsang
and Whalley~\cite{pqcd} using NLO pQCD, argued to have `incompatible'
normalization by Aurenche et al~\cite{aur2}, and believed to require
an intrinsic $<k_T>$ for partons as well as Cronin broadening of the
intrinsic momenta of partons~\cite{papp} leaves the field open to 
diverse interpretations.  From a purely empirical consideration also,
 it has been pointed out that~\cite{dks_scal}
the lowest energy $pp$ data for single photons are {\em not} consistent
 with the data at higher energies (and are too high).

In the light of the above, we take the view that the estimates of the
prompt photons given by Wong and Wang~\cite{wong}, give the
upper limit of these contributions.

Summarizing, we find that the single photons measured in the WA98 experiment
seem to confirm the formation of quark gluon plasma in the collision and
that  most of thermal radiation at higher transverse momenta
 seems to come from the annihilation of
quarks with scattering, which operates only if a plasma is formed.
As expected, the slope of the spectrum provides a very good measure
of the initial temperature reached in the collision.

This holds out the hope of a rich display of radiation of
photons from the quark matter at
RHIC and LHC energies in collisions involving heavy nuclei, as  much
larger temperatures are likely to be attained there. The long life of the
QGP phase at LHC energies will make it sensitive to such details like
the transverse flow (within the QGP phase itself!), which will be of immense
help in deciphering the properties of the quark-matter. 

{\em {Note added; further discussions}}

  We would like to take this opportunity to 
comment on some papers which have been posted~\cite{kam,jane,rus}
on the e-print archives {\em after} this paper was originally submitted.
This discussion is necessary in view of the important conclusions drawn
in this work, which differs in detail with findings in these papers.

The authors of Ref.\cite{kam} have used a very simple model to 
parametrize the evolution of the plasma. A spherical (!) expansion of the
plasma is envisaged which continues to radiate photons during the 
entire life-time at a {\em fixed} (average, effective) temperature. 
While this may be useful to suggest that there {\it is}  an additional
production of photons, this  approach is too simple to help us
arrive at quantities like initial temperature, $T_C$ etc.
Moreover as, at the relevant nucleon-nucleon energy 
($\sqrt{s}=$ 17.3 GeV) for the WA98 data under consideration, there is 
no pp data, these authors further scale the predictions of PYTHIA for
pp ( which required a $K$ factor of 3.2 and intrinsic parton momentum
for the E704  data at 19.4 GeV)
 to the WA98 data for $k_T >$ 2.5 GeV. This fore-closes any 
hope to get information about the origin of these photons by assigning them
to hard QCD interactions among partons whose distribution is obtained 
from structure function.

The authors of Ref.~\cite{jane} have used an early version of the
transverse expansion code used in the present work, which was 
originally adopted from the work of Ruuskanen and coworkers~\cite{hydro}. The
model uses a energy-density profile which is unform upto the
transverse radius $R$,  a hadronic gas which consists of only $\pi$,
$\rho$, $\omega$, $\eta$, and $a_1$ mesons and nucleons,
and uses the method of effective number of degrees of freedom 
at each temperature. This hadronic matter will have a much smaller
number of degrees of freedom at $T_C$, leading to a long lived mixed
phase compared to the case of rich hadronic matter used in the 
present work. The overall life time of the system will then be larger,
considerably enhancing the yields from the hadronic and the mixed phases.
Thus one would need only a smaller contribution from the QGP phase
to explain the data, as reported by these authors. It remains to be
seen, how these results will behave when the a correction for the
numerical factor of 4~\cite{MT} in rates given by Ref.~\cite{pat}
is made.
Further, the method of temperature dependent  number of degrees of freedom
will lead to a speed of sound $c_s=1/\sqrt{3}$ at all temperatures. A
uniform energy density profile does not reflect the actual situation either,
as it would follow wounded-nucleon distribution used in the present
work.  

Another aspect of this work is introduction of an {\em initial}
transverse velocity. It is well known from the pioneering work of
authors of Ref.~\cite{hydro} that the $<p_T>$ of the produced
particles can be arbitrarily increased if a strong initial transverse
flow is assumed. A look at Fig.3 of the present work also
suggests that one can use a larger formation time (smaller initial 
temperature) and an initial flow to {\em arbitrarily} increase the
large $k_T$ production. The authors of Ref.~\cite{rus} have also
introduced a large initial transverse velocity. 

However, it is known from the arguments of Ref.~\cite{hydro} that
in a head-on collisions of nuclei, introduction of initial transverse
velocity is not physically justified and one expects that, with the
exception of the outer surface, the produced matter would be
transversely at rest. At the most one may expect that the initial flow
may be stronger near the surface, which expands against the
vacuum, there is no conceivable mechanism to provide a significant
initial transverse collective motion, across the fluid.

The initial scattering among partons
will produce quarks and gluons pointing in random directions. In any
given volume element their momenta would be uniformly distributed in 
all directions, and rescatterings will then evolve a temperature and
pressure. This temperature and pressure (gradient) will initiate a
flow when the plasma starts expanding against the vacuum. 

Both these works, Ref.~\cite{jane,rus}, also include the
LO pQCD  predictions for the hard photons. While the authors of Ref.~\cite{jane}
include the intrinsic $k_T$ as in the work of Ref.\cite{wong}, they 
apparently they do not use a $K$-factor, though they use the LO pQCD.
 The authors of Ref.~\cite{rus} 
use a $K$-factor of 2 and find that they under-estimate the pp data
at  19 GeV by a factor of 7 (implying an effective $K$-factor
of $\sim$ 14 (!) over the LO prediction).

We have commented that the initial conditions deduced here provide
a quantitative description to the intermediate mass dilepton spectra
measured by the NA50 group~\cite{had_dil}. It is of interest to understand 
the origin of the differences in the initial conditions inferred by us
and those by the authors of Ref.~\cite{rapp}, who report an initial
temperature of about 200 MeV at an initial time of $\sim$ 1 fm/$c$.
 The fire-ball model used
in  Ref.~\cite{rapp}, envisages a cylinder  whose length and radius 
increase with time. The cylinder is assumed to be uniformly filled with
 plasma having temperature $T(t)$. We know for sure that the profile of the
energy density produced in such nuclear collisions can  {\em not} be uniform,
and this leads to additional gradients in the hydrodynamic evolution.
The model also does not account for the fact that the speed of sound is
large during the QGP phase, vanishingly small during the mixed-phase,
and varying with temperature during the hadronic phase if a rich 
hadronic equation of state is used. Thus, for example, during the mixed
phase their parameters $a_z$ and $a_T$ (which correspond to 
acceleration of the expanding surface) {\em must} vanish. 
 Even though the parameters of the model are
adjusted to give a transverse velocity equal to that deduced from
particle spectra and a transverse size deduced from interferometry, it
can not be expected to adequately reflect the rich history of evolution
of the plasma formed in nuclear collisions, and by 
extension, the initial conditions. 

These simplifying assumptions provide that the contribution of the
 hadrons to radiations from the system is larger, necessitating
only a small contribution from the QGP phase (and smaller initial temperature).

{\em  {Acknowledgments}}: 
{\sl We thank Patrick Aurenche, Terry Awes, Jean Cleymans, Joe Kapusta,  
Berndt M\"{u}ller, and Itzhak Tserruya for valuable comments.}

\begin{figure}
\epsfxsize=3.25in
\epsfbox{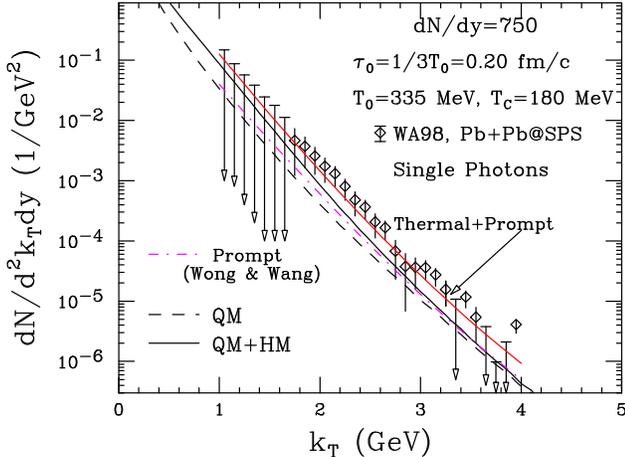}
\caption{
 Single photon production in $Pb+Pb$ collision at the CERN SPS.
A chemically and thermally equilibrated quark-gluon plasma is assumed
to be formed at $\tau_0=1/3T_0$ which expands, cools, enters into a
mixed phase and undergoes freeze-out from a hadronic phase.
 QM stands for radiations from the
quark matter in the QGP phase and the mixed phase. HM, likewise denotes
the radiation from the hadronic matter in the mixed phase and the
hadronic phase. 
Prompt photons are estimated
using NLO pQCD with the inclusion of
intrinsic \protect$k_T$ of partons (Wong and Wang~\protect\cite{wong}).
The (tail) ends of the arrows denote the upper limit of the production
at 90\% confidence limit.
}
\end{figure}

\begin{figure}
\epsfxsize=3.25in
\epsfbox{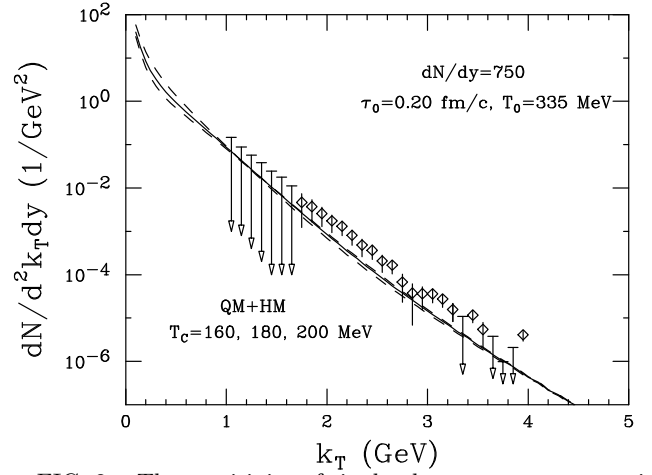}
\caption{
The sensitivity of single photon spectrum to critical
temperature. The solid curve is for $T_C=$ 180 MeV, while the
upper (lower) dashed curve is for 160 (200) MeV.
}
\end{figure}

\begin{figure}
\epsfxsize=3.25in
\epsfbox{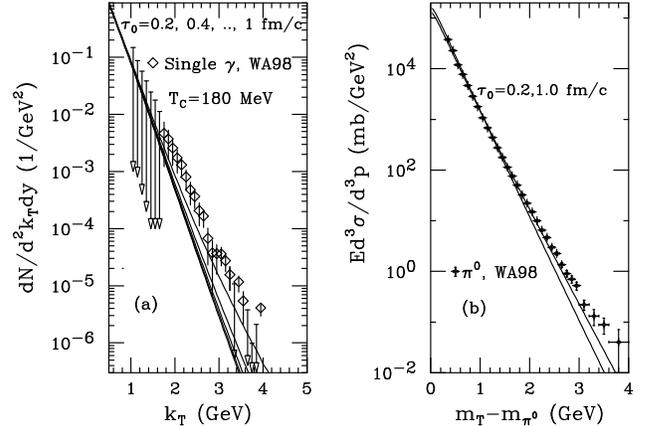}
\caption{
 The sensitivity of single photon (a) and pion 
 spectrum~\protect\cite{wa98pi} (b) to initial
time (temperature). The curves, from top to bottom, correspond to
initial times of 0.2, 0.4, 0.6, 0.8, and 1.0 fm/$c$ for (a) and to
0.2 and 1.0 fm/$c$ for (b).
}
\end{figure}


\begin{references}

\bibitem{early} E. L. Feinberg, Nuovo Cim. A {\bf 34}, 391 (1976);
                E. V. Shuryak, Phys. Lett. B {\bf 78}, 150 (1978).

\bibitem{qm99} See e.g.,{\sl Proc. Quark Matter '99},
         Nucl. Phys. A {\bf 661} (1999);
U. Heinz and M. Jacob, e-Print nucl-th/0002042;
 {\sl Physics and Astrophysics of Quark Gluon Plasma, (Proc. ICPA-QGP'97),
 Ed. B. C. Sinha, D. K. Srivastava, and Y. P. Viyogi}, (Narosa Publishing
 House, New Delhi, 1998).

\bibitem{wa98} M. M. Aggarwal et al., WA98 Collaboration,
 nucl-ex/0006007, Phys. Rev. Lett. 85, 3595 (2000).

\bibitem{wa80} R. Albrecht et al., WA80 Collaboration, Phys. Rev. Lett.
{\bf 76}, 3506 (1996).

\bibitem{prl} 
D. K. Srivastava and B. Sinha, Phys. Rev. Lett. {\bf 73}, 2421 (1994). 

\bibitem{others} See e.g.,
 N. Arbex et al., Phys. Lett. B {\bf 354}, 307 (1995);
A. Dumitru et al.  Phys. Rev. C {\bf 51}, 307 (1995); 
J. J. Neumann, D. Seibert, and G. Fai, Phys. Rev. C {\bf 51}, 1460 (1995);
J. Sollfrank et al., Phys. Rev. C {\bf 55}, 392 (1997).

\bibitem{zpc1} D. K. Srivastava and B. Sinha, 
Eur. Phys. Jour. C {\bf 12}, 109 (2000); Erratum  nucl-th/0103022.

\bibitem{crs}
J. Cleymans, K. Redlich, D. K. Srivastava,  
Phys. Rev. C {\bf 55}, 1431 (1997);
J. Cleymans, K. Redlich, D. K. Srivastava;
Phys. Lett. B {\bf 420}, 261 (1998). 

\bibitem{johanna}  See e.g., P. Braun-Munzinger, I. Heppe,  and J. Stachel 
 Phys. Lett. B {\bf 465}, 15 (1999);
 F. Becattini, J. Cleymans, A. Keranen, E Suhonen, and
 K. Redlich, e-Print hep-ph/0002267. 

\bibitem{had_dil} I. Kvasnikova, C. Gale, and D. K. Srivastava,
to be published; C. Gale, talk given at 30th Int. Symp. on Multiparticle
Dynamics, October 2000, Lake Balaton, Hungary, hep-ph/0102214;
 D. K. Srivastava, B. Sinha, I. Kvasnikova, and Charles Gale,
talk given at QM 2001, nucl-th/010301.

\bibitem{pat} P. Aurenche, F. Gelis, R. Kobes, and H. Zaraket,
Phys. Rev. D. {\bf 58}, 085003 (1998).

\bibitem{MT} D. Dutta et al. and P. Aurenche et al. Private Communication,
F. D. Steffen and M. H. Thoma, hep-ph/0103044.


\bibitem{joe} J. Kapusta, P. Lichard, and D. Seibert, Phys. Rev. D
{\bf 44}, 2774 (1991); Erratum, ibid D {\bf 47}, 4171 (1993).

\bibitem{rolf} R. Baier et al., Z. Phys. C {\bf 53}, 433 (1992).

\bibitem{zpc2} D. K. Srivastava, Eur. Phys. Jour. C {\bf 10}, 487 (1999);
Erratum nucl-th/0103023.

\bibitem{pcm}
 K. Geiger and B. M\"{u}ller, Nucl. Phys. B {\bf 369}, 600 (1992);
K. Geiger, Phys. Rep. {\bf 258}, 376 (1995).

\bibitem{pcmphot} D. K. Srivastava and K. Geiger, Phys. Rev. C {\bf 58},
                  1734 (1998).

\bibitem{wong} C. Y. Wong and H. Wang, Phys.  Rev. C {\bf 58}, 376 (1998)

\bibitem{bj} J. D. Bjorken, Phys. Rev. D {\bf 27}, 140 (1983);
              R. C. Hwa and K. Kajantie, Phys. Rev. D {\bf 32}, 1109 (1985).

\bibitem{kms} J. Kapusta, L. McLerran, and D. K. Srivastava, Phys. Lett.
             B {\bf 283}, 145 (1992);
 D. K. Srivastava et al., Phys. Lett.  B {\bf 283}, 285 (1992);
 D. K. Srivastava and B. Sinha, J. Phys. G {\bf 18}, 1467;
 R. Vogt et al., Phys. Rev. D {\bf 49}, 3345 (1994); 
S. Gavin et al., Phys. Rev. C {\bf 54},2606 (1996).

\bibitem{kars} F. Karsch,  
Plenary talk at the 17th International Symposium on Lattice Field Theory
 (LATTICE 99), Pisa, Italy, e-Print  hep-lat/9909006.

\bibitem{li} Li Xiong, E. Shuryak, and G. E. Brown, Phys. Rev. D
{\bf 46}, 3798 (1992).

\bibitem{hydro} H. von Gersdorff et al., Phys. Rev. D {\bf 34}, 794 (1986);
 P. V. Ruuskanen, Acta Phys. Pol. B {\bf 18}, 551 (1986).
 K. Kajantie, M. Kataja, L. McLerran, and P. V. Ruuskanen, Phys. Rev. D
{\bf 34}, 811 (1986).

\bibitem{kap} See e.g., J. Kapusta and A. Mekjian, Phys. Rev. D {\bf 33},
                        1304 (1986);
T. Matsui, B. Svetitsky, and L. McLerran, Phys. Rev. D {\bf 34}, 783
(1986), ibid {\bf 34}, 2047 (1986);
P. Koch, B. M\"{u}ller, and J. Rafelski, Phys. Rep. {\bf 4}, 167
(1986).

\bibitem{pcmphot2} D. K. Srivastava, {\sl RHIC Physics and Beyond, Proc.
Kay Kay Gee Day, Ed. B. M\"{u}ller and R. D. Pisarski},
(American Institute of Physics, New York, 1999) p. 142.


\bibitem{wa98pi} M. M. Aggarwal et al. Phys. Rev. Lett. {\bf 81},
4087 (1998); Erratum ibid {\bf 84}, 578 (2000).

\bibitem{xu} F. Wang and N. Xu, Phys. Rev. C {\bf 61}, 021904(R) (2000).

\bibitem{sat} K. J. Eskola, K. Kajantie, P. V. Ruuskanen, and K. Tuominen,
Nucl. Phys. B {\bf 570}, 379 (2000); see also H. Satz, Nucl. Phys. A {\bf 661},
104c (1999).

\bibitem{kam} K. Gallmeister, B. K\"{a}mpfer, and O. P. Pavlenko,
hep-ph/000613, Phys. Rev. C {\bf 62}, 057901 (2000). 

\bibitem{jane} J. Alam, S. Sarkar, T. Hatsuda, T. K. Nayak,
and B. Sinha, Phys. Rev. C {\bf 63}, 021901 (2001), hep-ph/0008074.

\bibitem{rus} D. Y. Peressounko and Y. E. Pokrovsky, hep-ph/0009025.

\bibitem{owens} J. F. Owens, Rev. Mod. Phys. {\bf 59}, 465 (1987).

\bibitem{pqcd} W. Vogelsang and M. R. Whalley, J. Phys. G. {\bf 23},
                 A1 (1997). 

\bibitem{aur2} P.Aurenche, M. Fontannaz, J. Ph. Guillet, B. Kniehl,
E. Pilon, and M. Werlen, Eur. Phys. J. C {\bf 9}, 107 (1999).

\bibitem{papp} G. Papp, P. Levai, and G. Fai, Phys. Rev. C {\bf 61}, 021902
(2000).

\bibitem{dks_scal} D. K. Srivastava, nucl-th/0102005.

\bibitem{rapp} R. Rapp and E. Shuryak, Phys. Lett. B {\bf 473}, 13 (2000).

\end{references}
\end{document}